\documentclass[]{aa}
\usepackage{times}
\usepackage{natbib}
\usepackage[dvips]{graphicx}
\usepackage{psfig,longtable,lscape}
\usepackage{psfig}
\usepackage{rotating}
\bibpunct{(}{)}{;}{a}{}{,}

\def\Swift{{\em Swift}}
\def\1E{1E1207.4-5209}

\def\ltsima{$\; \buildrel < \over \sim \;$}
\def\simlt{\lower.5ex\hbox{\ltsima}}
\def\gtsima{$\; \buildrel > \over \sim \;$}
\def\simgt{\lower.5ex\hbox{\gtsima}}

\begin{document}

\title{Using GRO J1655-40 to test \Swift/BAT as a monitor for bright hard X-ray sources}

\author{F. Senziani\inst{1,2,3}, G. Novara\inst{1,2}, A. De Luca\inst{2},
P. A. Caraveo\inst{2}, T.Belloni\inst{4}, G. F. Bignami\inst{5,6}}

\institute{Universit\`a di Pavia, Dipartimento di Fisica Teorica e Nucleare, Via Ugo Bassi 6, I--27100 Pavia (I)
         \and INAF-Istituto di Astrofisica Spaziale e Fisica Cosmica, Via Bassini 15, I--20133 Milano (I)
     \and Universit\'e Paul Sabatier - 118, Route de Narbonne - 31062 Toulouse (F)
     \and INAF-Osservatorio Astronomico di Brera, Via E.Bianchi 46, I-23807 Merate (LC), Italy.
     \and Agenzia Spaziale Italiana, Via Liegi 26, I-00198 Roma, Italy
     \and Istituto Universitario di Studi Superiori (IUSS), c/o Collegio Giasone del Maino, Via Luino 4, I--27100 Pavia (I)
}

\offprints{Fabio Senziani, senziani@iasf-milano.inaf.it}

\date{}

\authorrunning{F. Senziani et al.}

\titlerunning{BAT as a bright source monitor}

\abstract {While waiting for new gamma-ray burst detections, the
Burst Alert Telescope (BAT) onboard Swift covers each day
$\sim$50\% of the sky in the hard X-ray band (``Survey data'').
The large field of view (FOV), high sensitivity and good angular
resolution make BAT a potentially powerful all-sky hard X-ray
monitor, provided that mask--related systematics can be properly
accounted for.} {We have developed and tested a complete procedure
entirely based on public Swift/BAT software tools
to analyse BAT Survey data, aimed at assessing the flux and
spectral variability of bright sources in the 15-150 keV energy
range.} {Detailed tests of the capabilities of our procedure
were performed
focusing, in particular, on the
reliability of spectral measurements over the entire BAT FOV.
First, we analyzed a large set of Crab observations, spread over
$\sim$7 months. Next, we studied the case of GRO J1655-40, a
strongly variable source, which experienced a 9-month long
outburst, beginning on February 2005. Such an outburst was
systematically monitored with the well-calibrated PCA and HEXTE
instruments onboard the RXTE mission. Thanks to the good BAT
temporal coverage of the source, we have been able to cross-check
BAT light-curves with simultaneous HEXTE ones.} 
{The Crab tests
have shown that
our procedure
recovers both the flux and the
source spectral shape over the whole FOV of the BAT instrument.
Moreover, by cross-checking GRO J1655-40 light-curves obtained by
BAT and HEXTE, we found the spectral and flux evolution of the
outburst to be in very good
agreement. 
Using our procedure, BAT reproduces HEXTE
fluxes within a 10-15\% uncertainty with a $3\sigma$ sensitivity of $\sim20$ mCrab for an on-axis source, thus establishing its
capability to monitor the evolution of relatively bright hard
X-rays sources. }
{}


\maketitle

\section{Introduction}

The Burst Alert Telescope (BAT; \citep{Barthelmy}), the gamma-ray
instrument on board \Swift~\citep{Gehrels}is a highly sensitive
coded mask instrument optimized in the 15-150 keV energy range.
Its large ($\sim$ 2 sr) D-shaped FOV allows for the coverage
of $\sim$ 1/6 of the sky in a single pointing. BAT was designed
to be an efficient detector for Gamma-Ray Bursts (GRBs) and is now
fulfilling its pre-launch expectations, delivering $\sim$ 100
GRBs/yr \footnote{{\em http://swift.gsfc.nasa.gov/docs/swift/bursts/index.html}}.

While waiting for new GRBs, BAT collects a huge amount of data on
the hard X-ray sky. Indeed, a sensitive all-sky survey in the
15-150 keV energy range will be one of the major outcomes of the
\Swift~mission. With an expected limiting flux of $\sim0.2$ mCrab
at high Galactic latitude and $\sim3$ mCrab at low Galactic
latitude (with a 4-year dataset), the Swift survey is expected to
be $\sim$10 times deeper than the HEAO1 A4 reference all-sky hard
X-ray survey \citep{Levine_1984}, performed more than 25 years ago. Preliminary
results, based on 3 months of data, have been published by
\citet{Markwardt_2005}.  

Owing to its large FOV, high sensitivity and good angular
resolution, BAT could perform an efficient monitoring of
high-energy sources. Indeed, count-rate light-curves (in the
15-50 keV energy range) for more than 400 known sources, updated on a
single orbit basis, have been recently made available on the web
in the ``BAT Hard X-ray transient monitor'' facility of the
Goddard Space Flight Center \footnote{{\em http://swift.gsfc.nasa.gov/docs/swift/results/transients/}} \citep{Krimm_2006}. 

To fully exploit the data collected by BAT, however, 
count rate light curves should be converted into flux light curves.
This would allow a continuous assessment of the varying 
sources' spectral shape. Thus, we have developed a procedure aimed
at extracting spectral information for bright hard X-ray sources falling
into the BAT field of view.

Tests were made to verify the reliability of the results of our method, 
focusing on the spectral performances of the BAT instrument.
First we analyzed a large number of observations of the Crab
nebula (and its pulsar), the classical calibration source for
X-ray instruments, performed under different observing conditions.
Next, we performed a detailed study of the 9-month long 2005
outburst of the galactic microquasar GRO J1655-40. Such event was
also carefully monitored with the narrow field instruments PCA and
HEXTE on-board the Rossi X-Ray Timing Explorer (RXTE), making it
possible to cross-check the BAT results with those obtained
simultaneously by an independent well calibrated instrument.

The paper is organized as follows. After a brief overview of BAT
survey data structure (section \ref{Swift_BAT_Survey_Data}), we
outline our automatic data analysis pipeline. The tests performed
on the Crab are reported in section \ref{Crab_calibration},
focusing on the reliability of spectral 
results
as a function
of the source position within the BAT FOV. The requirements to
combine different BAT datasets are spelled out in section
\ref{Delta_flusso_delta_offset}. 

After assessing the reliability of our pipeline using the very bright and
steady Crab as a reference source, we went on to study the 2005
outburst of GRO J1655-40. 
Our  RXTE
spectral analysis is described in section \ref{RXTE_analysis}.
Finally, we compare BAT and RXTE spectral results, thus assessing
the performances of our pipeline in order to use the BAT data to monitor the behaviour of a strongly variable
source (section \ref{GRO_results}).
Appendices A and B provide technical details on the 
data analysis pipelines.

\section{\Swift/BAT Survey Data}\label{Swift_BAT_Survey_Data}

BAT is a coded mask instrument with a detecting area of 5200
cm$^2$. The detector is an array of 32768 CdZnTe elements
operating in the 15-150 keV energy band with a good sensitivity
and energy resolution \citep{Barthelmy}. The BAT instrument is
operated in photon-counting mode. Photons interacting with the
detector are processed (events) and then are tagged with an
associated time of arrival, detector number and energy. Such
information is stored on-board in a memory buffer 
which may contain
$\sim$10 minutes of data (depending on the actual count rate).
Data are analyzed in real time using several algorithms in order
to detect new GRBs. In case of a GRB trigger, the event buffer is
sent to the ground for more detailed analysis (event files),
otherwise it is organized on board in a Detector Plane Histogram
(DPH) and then sent to the ground. DPHs are three
dimensional histograms: for a given buffer set, every cell
contains the number of events received in one of 32768 pixels of the detector plane and in one of 80 energy channels.
Such histograms are accumulated over a typical 5 minutes time
interval and then stacked as independent rows in a ``Survey''
data file.
Together with auxiliary files (which contain all spacecraft-related
 information
for a given observation), they are the standard basic
products for BAT non-GRB science.

All \Swift~data and software are publicly available and they have been downloaded at Heasarc-U.S. web site \footnote{{\em http://heasarc.gsfc.nasa.gov/cgi-bin/W3Browse/swift.pl}}.
Version 2.4 of the \Swift~software \footnote{{\em http://swift.gsfc.nasa.gov/docs/software/lheasoft}} was used for data reduction.

\section{BAT as a monitor: calibration with the Crab}\label{Crab_calibration}

\begin{figure}
\begin{center}
     \hspace{0cm}\psfig{figure=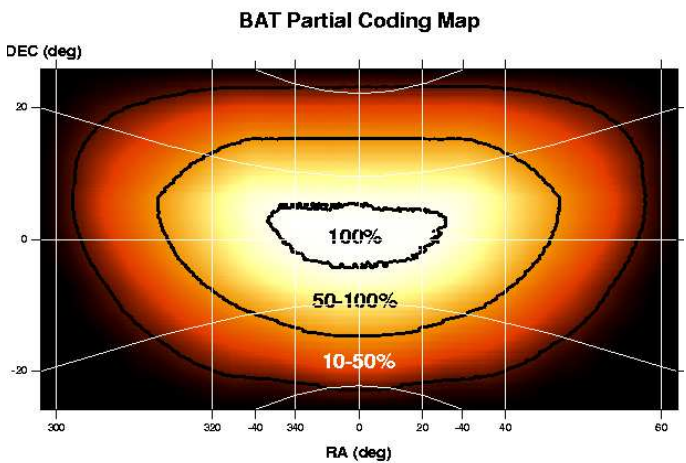,width=9cm}
      \caption[]{The D-shaped BAT Field of View. Fully coded (100\%), half-coded (50\%) down to the 10\% coded fraction contours are indicated.}
      \label{BAT_FOV_coded_fraction}
\end{center}
\end{figure}

BAT has a very large field of view.  In order to fully exploit
such a capability - which is crucial to use 
BAT as a monitor for the hard X-ray sky - as a first step we
have to assess the stability of the source flux as reconstructed
by our procedure as a function of the source position within
the FOV. Indeed, different positions within
the instrument FOV correspond to different mask coded fractions
(see Figure ~\ref{BAT_FOV_coded_fraction}).
To this aim, we will study a large sample of observations
of the bright and steady Crab, the source used to calibrate
the BAT spectral response. This will be a crucial test to assess
the capability of our method  to extract spectral and flux information
across the whole instrument FOV. 

We implemented an automated data analysis pipeline
based on a single-DPH row logic, i.e. extracting products for each
single DPH row, following three steps.

\begin{enumerate}

\item ``Good'' data are carefully selected. Each DPH is checked for
the overall count rate, pointing stability, occultations of part
of the FOV by the Earth or the Moon, as well as for other possible
sources of systematic effects (see Appendix
\ref{Preliminary_data_selection_and_preparation} for details).

\item Selected DPHs are processed to compute the source flux. 

To this aim we used the so called ``Mask-weighting'' technique. Such an approach is based on the ``mask-weight map'',  accounting for the fraction of the detector which is
shadowed by the mask with respect to the source position. Applying such a mask to the DPH yields a background-subtracted
spectrum of the source (for a detailed description, see Appendix
\ref{Mask_weighting}).

\item We also investigated the possibility to combine
different DPHs, in order to increase the exposure time (and thus
the source signal to noise) beyond the typical $\sim5$ min
integration of a single DPH. A careful check of the effects of DPH
pointing offsets was performed, in order to assess the maximum
``misjudgement'' allowed to minimize flux loss in the combined
DPH.

\end{enumerate}

All the details of our pipeline are given in Appendix \ref{Crab_pipeline}
(to study the Crab, we use steps 1,2 and 3 described there).

\subsection{Data selection}\label{Crab_data_preparation}

Our Crab data set encompasses all observations, with the source
within the BAT FOV, collected between 2005/01/01 and 2005/06/30.
Such a sample includes 365 observations, including a total of 4626
DPH rows (for a total observing time of $\sim$1.46 Ms).
We note that such a data set is larger than the one presented on 
the public BAT Digest pages\footnote{\em http://heasarc.nasa.gov/docs/swift/analysis/bat\_digest.html} 
describing details on the instrument calibration (44 grid locations in the BAT FOV).
After the data screening stage (Appendix \ref{Preliminary_data_selection_and_preparation}), 1014 rows ($\sim$ 20\%) were
rejected. Details on the impact of each specific filtering criterium
are given in Table \ref{Table_filter}.

\begin{table}[tbp]
\begin{center}
\footnotesize{
\begin{tabular}{lc} \hline
Detectors not enabled & 36 \\
Earth contamination + SAA & 121 \\
Star tracker unlocked & 264 \\
Pointing unstable & 317 \\
High count rate & 328 \\
Earth/Moon/Sun source occultation & 297 \\
\hline \\
\end{tabular}}
\caption{Number of DPH rows of our Crab sample rejected after the filtering stage.}
\label{Table_filter}
\end{center}
\end{table}

Moreover, 34 rows ($<1\%$) could not be processed owing to the
lack of the relevant housekeeping files. We are then left with  $\sim3600$ good DPH rows which were used for image, as well as for spectral, analysis.

Of course, the Crab position varied greatly within the BAT FOV.
Table~\ref{Table_histog_deg} provides the statistic of the source
position in the good BAT data set.

\begin{table}[tbp]
\begin{center}
\footnotesize{
\begin{tabular}{ll} \hline
Deg off-axis & Processed rows \\
\hline
0-5 & 306 \\
5-10 & 0 \\
10-15 & 0 \\
15-20 & 108 \\
20-25 & 33 \\
25-30 & 299 \\
30-35 & 908 \\
35-40 & 253 \\
40-45 & 816 \\
45-50 & 295 \\
50-55 & 323 \\
55-60 & 207 \\
60-.. & 0 \\
\hline \\
\end{tabular}}
\caption{Distribution of the Crab location in the FOV for the
processed sample of DPH rows. The source position is given in
degrees from the center of the BAT FOV.} \label{Table_histog_deg}
\end{center}
\end{table}

\subsection{Results I: flux evaluation }\label{Crab_results}

BAT has a coded aperture imaging system. The net count-rate of a
source at a given position in the sky may be extracted using the
mask-weighting technique\footnote{\em http://swift.gsfc.nasa.gov/docs/swift/analysis/swiftbat.pdf}. Basically, it consists in assigning to
each detection element a weight (from -1 to +1), depending on the
fraction of the detector which is shadowed by the mask with
respect to the source position. By applying such a weight map to
a DPH, it is then possible to extract a background-subtracted
spectrum for a source of known position (see Appendix \ref{Mask_weighting}
for details).

\begin{figure}[htbp]
\begin{center}
      \hspace{0cm}\psfig{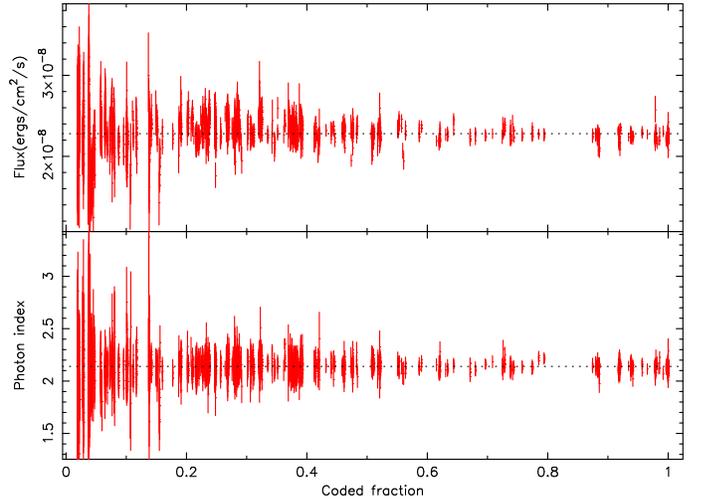}
      \caption[]{Crab flux in the 10-100 keV energy band (upper panel) and spectral
      photon index (lower panel) as a function of the coded fraction, as estimated with
      mask-weighting technique using celestial Crab coordinates.
      The horizontal dotted lines represent the best fit to the constant model (see table \ref{Table_Crab_stat} for details).}
      \label{fig_flux_phindx_pcod_cat}
\end{center}
\end{figure}

The Crab mask-weighed spectrum was extracted using the nominal coordinates.
All Crab spectra were fitted with a power law in the 10-100 keV
range. Best fit photon indices and normalization factors, as well
as observed fluxes, were computed.
A plot of the Crab 10-100 keV flux in physical units (erg
cm$^{-2}$ s$^{-1}$) as a function of partial coding fraction is
shown in Fig.~\ref{fig_flux_phindx_pcod_cat} (upper panel). A
similar plot with the values of the photon index is shown in the
lower panel of the same figure.
Both the flux and the photon index are in good agreement
with the values assumed for the BAT instrument calibration (see BAT Digest pages).
Both the flux and photon index values are very
stable as a function of the source location within the BAT FOV.
As shown in Table \ref{Table_Crab_stat}, the spread of the values
is within 8.1\%, and is further reduced to 5.9\%, if only coded
fraction $>$0.1 are selected.

\begin{figure}
\begin{center}
       \hspace{0cm}\psfig{figure=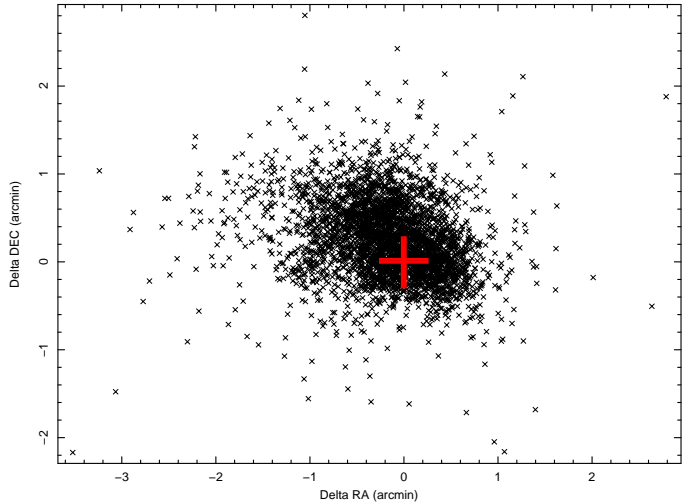,width=9cm}
      \caption[]{Dispersion of Crab detection coordinates around the nominal source
      position (red cross).We note that the coordinates
dispersion is not perfectly centered on the Crab nominal
position. Probably this is related to systematic image centroid
shifts as a function of the position in the FOV (see BAT Digest pages), as well as to
the distribution of the Crab detector coordinates in our
dataset.}
      \label{fig_coord_diff}
\end{center}
\end{figure}

Then, we investigated whether the use of 
the target coordinates as derived from a source detection
(instead of the nominal ones)
could yield different results for the extraction of spectral information.
For each DPH, an image of the BAT FOV was produced, and standard
source detection was performed using publicly available Swift
analysis software tools (see Appendix \ref{Imaging_Analysis} for a detailed description).
The Crab was detected
in all but 39 cases. The target detection
coordinates agree well with the known Crab position and the
dispersion around the nominal source coordinates is very small, as
shown graphically in Figure \ref{fig_coord_diff} and numerically
in Table \ref{Table_coord_diff}.

\begin{table}[tbp]
\begin{center}
\footnotesize{
\begin{tabular}{c|cc|cc}
& \multicolumn{2}{c}{Detection Coordinates} & \multicolumn{2}{c}{Catalogue Coordinates} \\
\hline
& \multicolumn{4}{c}{\textbf{FLUX}} \\
\hline
Coded fraction & Mean & rms & Mean & rms \\  
\hline
$>$ 0 & 2.28E-08 & 7.7\% & 2.28E-08 & 8.1\% \\  
$>$ 0.1 & 2.28E-08 & 5.0\% & 2.28E-08 & 5.9\% \\  
\hline
& \multicolumn{4}{c}{\textbf{PHOTON INDEX}} \\
\hline
$>$ 0 & 2.14 & 0.09 & 2.14 & 0.11 \\   
$>$ 0.1 & 2.14 & 0.07 & 2.14 & 0.08 \\  
\hline
\end{tabular}}
\caption{Flux and power law statistic of Crab dataset. Fluxes in the 10-100 keV energy band are expressed in ergs cm$^{-2}$s$^{-1}$ units.}
\label{Table_Crab_stat}
\end{center}
\end{table}

\begin{table}[tbp]
\begin{center}
\footnotesize{
\begin{tabular}{l|cc}
& Coded fraction $>$ 0 & Coded fraction $>$ 0.1 \\
\hline
$< 1 arcmin$ & 84.8\% & 90.9\% \\
$< 2 arcmin$ & 98.6\% & 99.4\% \\
$< 3 arcmin$ & 99.9\% & 100.0\% \\
\hline
\end{tabular}}
\caption{Catalogue and detection coordinates percent difference.}
\label{Table_coord_diff}
\end{center}
\end{table}

\begin{figure}[htbp]
\begin{center}
    \hspace{0cm}\psfig{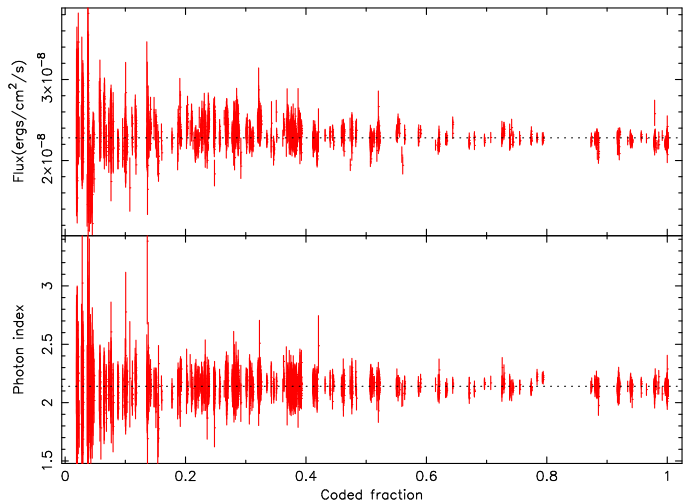}
      \caption[]{Crab flux in the 10-100 keV energy band (upper panel) and spectral
      photon index (lower panel) as a function of the coded fraction,
      as estimated with mask-weighting technique using Crab detection coordinates.
      The horizontal dotted lines represent the best fit to the constant model (see table \ref{Table_Crab_stat} for details).}
      \label{fig_flux_phindx_pcod_det}
\end{center}
\end{figure}

Then, we repeated the mask weighting, as well as the spectral analysis,
using as a starting point the source detection coordinates
instead of the nominal ones. Results were found to be virtually undistinguishable
(See Figure~\ref{fig_flux_phindx_pcod_det} and Table~\ref{Table_Crab_stat}).

\begin{figure}[htbp]
\begin{center}
       \hspace{0cm}\psfig{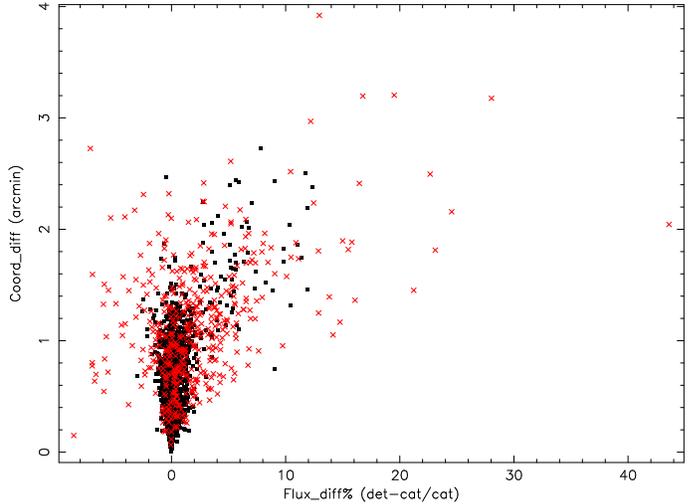}
      \caption[]{Flux difference using detection coordinates and nominal coordinates as a function of coordinate difference. Black squares represent the rows for which Crab
was located at coded fraction greater than 0.1, whereas red
crosses those for which the coded fraction was smaller or equal
to 0.1 . We note that the use of detection coordinates induces a
slight source flux overestimate which can be quantified in
$\sim$0.5 \%.}
      \label{fig_flux_coord_diff}
\end{center}
\end{figure}

Figure \ref{fig_flux_coord_diff} shows the flux difference as a
function of the offset between the real Crab coordinate (in red)
and those provided by the detection algorithm (in black).
Considering the totality of the Crab data, we note that 96.0\% of
the detections lie within 3 arcmin from the source celestial
coordinates and that their flux values differ by, at most,
$\pm$5\%. Considering only flux values extracted from observations
with coded fraction greater than 0.1, the $\pm$5\% threshold is
met in 98.5\% of the cases.

\subsection{Results II: stacking multiple DPHs}\label{Delta_flusso_delta_offset}

DPH have a typical integration time of 5 minutes. A single-DPH
analysis may be desirable in order to maximize time resolution.
However, especially when dealing with faint sources, it may be
important to increase statistics, at the expense of time
resolution, by merging several DPHs. In general, different DPHs
are collected with different mean satellite attitudes. Such a
difference must be carefully checked before proceeding with the
merging, since any offset may cause a partial loss
in the reconstructed source flux.
In order to quantify such an effect as a function of the pointing
offset, we performed a simple test with the Crab. We started by
selecting, among the data set described in section \ref{Crab_data_preparation}, DPHs where the Crab happened to be in three
different FOV regions (at coded fraction 1, $\sim$0.5,
$\sim$0.1). For each region, we selected a reference DPH 
together with a group of nearly co-aligned DPHs, having
ROLL angles within 1 arcmin of the reference one,
as well as pointing offsets (in R.A. and Dec.) within 7 arcmin of the
reference one.

Then, each of the selected DPH rows was summed to the
reference one, obtaining stacked DPH pairs, each 
being characterized by a ``pointing offset'' ranging 
from 0 to 7 arcmin.
For each of such stacked pairs, we performed a
spectral analysis and we extracted the Crab flux,
to be compared with the one obtained from the reference DPH. 
Details on the handling of different attitude files
for the construction and analysis of a pair are given in Appendix \ref{Aspect}.

The
resulting flux-losses 
with respect to the reference DPH, as a function of the offset, for different
coded fractions, are given in Table \ref{table_offset}.

Although our investigation is far from complete, results
suggest that significant flux losses ($>$ 5\%) may occur,
especially for target position at low coded fractions,
when stacking different DPHs with a pointing offset
larger than 2 arcmin.
Thus, 
we decided conservatively
to stack DPHs only if their pointings are within 1.5
arcmin.

\begin{table}[tbp]
\begin{center}
\footnotesize{
\begin{tabular}{ccc}
Offset (arcmin)& Flux loss \\
\hline
\multicolumn{2}{c}{\textbf{Pcode = 1}} \\
\hline
1   & 0.1\% \\
2   & 0.3\% \\
3   & 1.9\% \\
5   & 5.7\%  \\
7   & 11.1\% \\
\hline
\multicolumn{2}{c}{\textbf{Pcode = 0.5}} \\
\hline
1   & 1.2\% \\
2   & 3.6\% \\
2.7 & 5.2\%  \\
4   & 19.4\% \\
\hline
\multicolumn{2}{c}{\textbf{Pcode = 0.1}} \\
\hline
2   & 7.7\% \\
\hline
\end{tabular}}
\caption{Crab flux loss for two DPH stacked as a function of the DPH pointing offset. Different positions of the Crab in the FOV have been considered.}
\label{table_offset}
\end{center}
\end{table}

\section{Monitoring a strongly-variable source: the case of GRO J1655-40}\label{GRO}

Having assessed the overall correctness and reliability 
of our procedure on the bright Crab, we will proceed with
the study of
the microquasar GRO J1655-40, a source
definitely fainter than the Crab and one known to be strongly
variable, both in flux and in spectral shape.

GRO J1655-40 had a large outburst in 2005. Such an event
started in the middle of February
(it was discovered on February 17.99 during
Galactic bulge scans with the RXTE/PCA instrument \citep{Markwardt_2005_GRO})
and lasted for more than 9 months. In what follows we shall take
advantage of a very large database serendipitously collected by
the BAT instrument during the whole outburst event as well as of
the systematic monitoring performed by the RXTE satellite.

This will allow us to compare our BAT results with
quasi-simultaneous results obtained with the well calibrated
instruments on-board RXTE. Such a cross-check will yield a very
robust assessment of the capabilities of our analysis method as well as of the 
potentialities of BAT as a monitor for a
(relatively) bright, strongly variable source.

\subsection{BAT data analysis}\label{GRO_pipeline}
All BAT observations covering the field of GRO J1655-40,
collected between 2005/01/22 and 2005/11/11, were retrieved. The
complete dataset includes 796 observations, for a total of 8724
DPHs, corresponding to $\sim$2.6 Ms observing time.

All the data analysis was performed by our automatic pipeline. As
a first step, good data are selected, according to the
prescription described in Appendix
\ref{Preliminary_data_selection_and_preparation}. A total of 2080
($\sim$24\%) DPHs were discarded after data screening. Such a
percentage is compatible with that found for the Crab dataset in
section \ref{Crab_results}.

Next, well-aligned, contiguous DPHs are combined
up to a maximum integration time of
1 hour. As a result, we obtained 1650 merged DPHs.

Then, from each data block, a spectrum is extracted with the
mask-weighting technique, and the appropriate response matrix is
produced.

An automatic spectral analysis is then carried out in XSPEC. After evaluating the source signal-to-noise, spectra
with no signal (S/N=0)were discarded. This resulted in the
rejection of 378 spectra ($\sim$ 23\% of the total). Low S/N
spectra (with source detection below $\sim4\sigma$ level)
are used to set an upper limit to the source flux. Contiguous,
low-S/N spectra are summed, as well as their response matrices, in
an attempt to increase the statistics, and the spectral analysis
repeated on such combined spectra. High-S/N spectra are used for a
complete spectral fit using a power law model.

A detailed description of the data analysis pipeline is given in
Appendix \ref{Crab_pipeline}. The complete pipeline used for our study of GRO
J1655-40 consists of steps 1, 2, 4, and 5 described there.

\subsection{RXTE data analysis}\label{RXTE_analysis}

RXTE monitored the whole outburst of GRO J1655-40 since its
discovery \citep{Markwardt_2005_GRO}. The dataset is composed of
490 observations, performed between 2005-02-26 and 2005-11-11.
Each observation has a typical integration time of $\sim$1.5 ks,
for a total observing time of $\sim$664 ks.

Spectral data extracted from the complete dataset
have been kindly made available to the community by the MIT group \footnote{{\em http://tahti.mit.edu/opensource/1655/}}.
Spectra for both source and background as well as response matrices and effective
area files have been retrieved from their web site.

In order to ease a comparison of the results between different
instruments, the same choices of energy range and spectral model
were adopted. Thus, only data from the HEXTE instrument
(operating in the 20-200 keV energy range) were used for the
spectral fits. The spectral analysis was performed with an
automatic pipeline based on the same algorithm adopted for BAT
(as described in Appendix \ref{Crab_pipeline}).
We discarded 53 low quality spectra, with S/N$<$3.5.

As a further step, data from the PCA instrument (operating in the
2-60 keV energy range) were used to extract a simple light curve
(cts s$^{-1}$) in the soft (20-30 keV) energy range. An analogous
light curve for the hard range (30-100 keV) was also extracted
from HEXTE data and an hardness ratio plot was produced.

In addition, we downloaded public RXTE All Sky Monitor (ASM) data collected
during the whole GRO J1655-40 outburst and extracted a count rate
light curve in the 2-10 keV range.

\subsection{Results}\label{GRO_results}

\begin{figure*}[htbp]
\begin{center}
       \hspace{0cm}\psfig{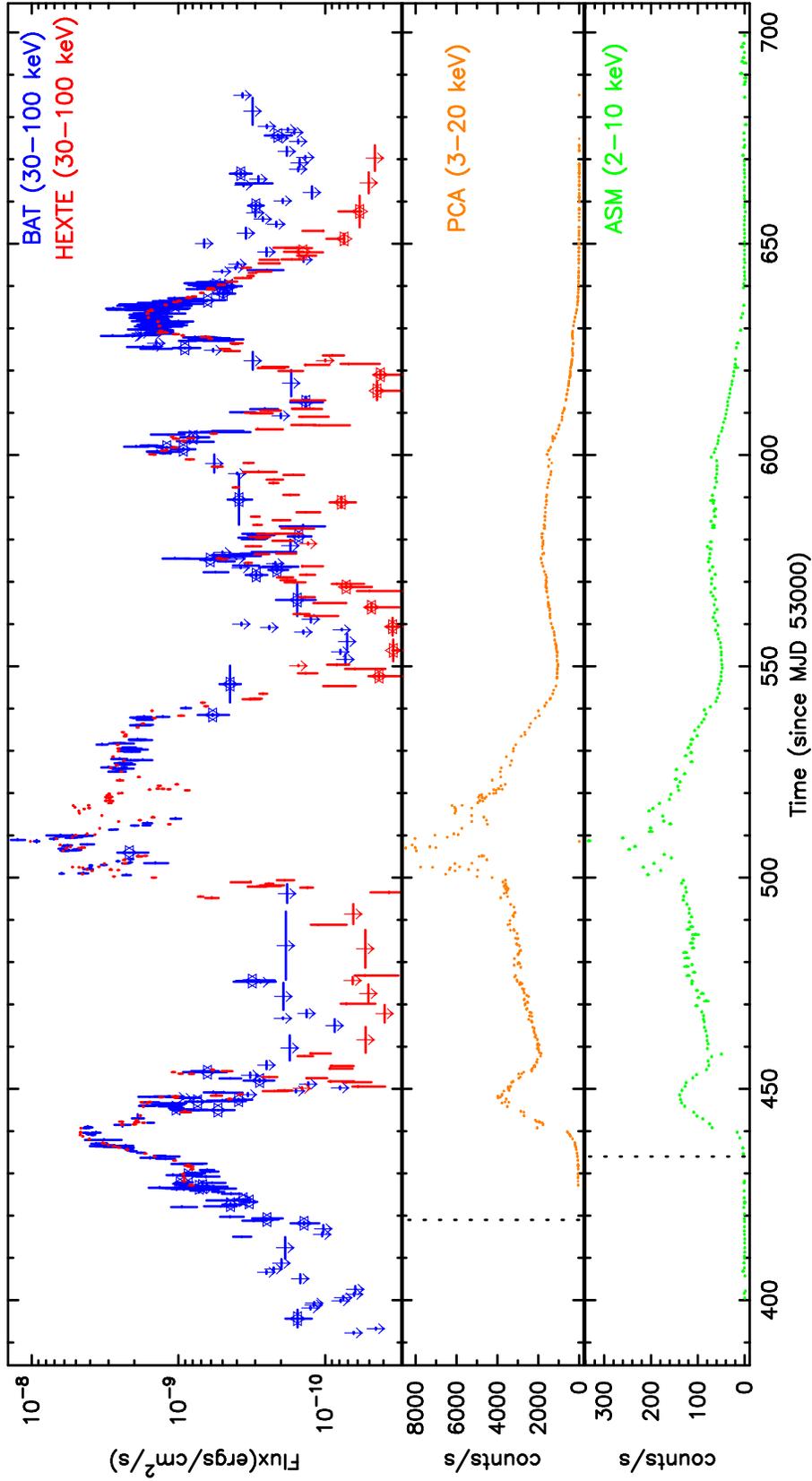}
      \caption[]{GRO J1655-40 complete 2005 outburst as seen from \Swift/BAT and RXTE/HEXTE
      (upper panel), RXTE/PCA (central panel), and RXTE/ASM (lower panel). I
      n the upper panel, stars mark BAT and HEXTE points corresponding to averaged spectra. 3$\sigma$ upper limits are also plotted, marked by arrows.
      Vertical dotted bars in the central and lower panels represent the time of the first detection of the outburst obtained with PCA and ASM respectively.}
      \label{fig_GRO_flux}
\end{center}
\end{figure*}

The complete light curve of the outburst of GRO J1655-40 as seen
by BAT (in erg cm$^{-2}$ s$^{-1}$) is shown in Figure
\ref{fig_GRO_flux} (top panel). HEXTE measurements are also shown,
to allow for a direct comparison. We plotted in the same figure
the light curves extracted from the PCA data (count rate in 3-20
keV, central panel) and from the ASM data (count rate in 2-10 keV,
bottom panel). Errors are at $1\sigma$ level.

\begin{figure*}[htbp]
\begin{center}
       \hspace{0cm}\psfig{figure=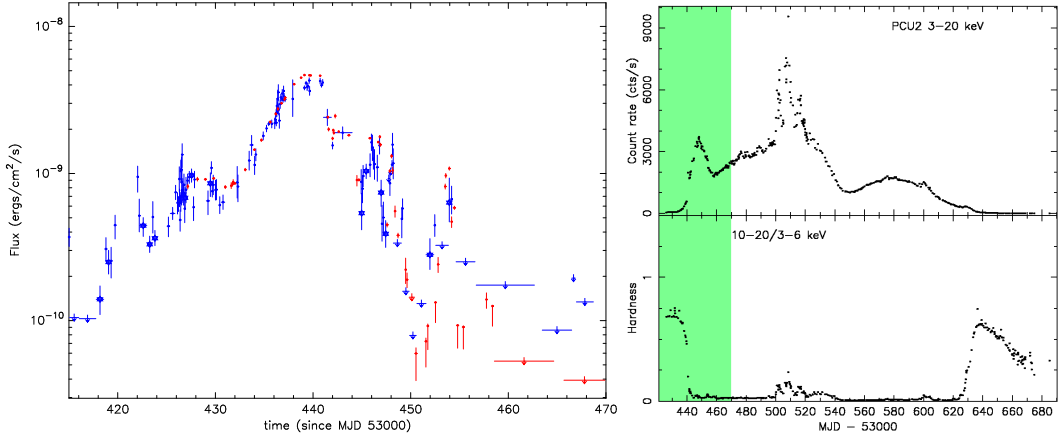,width=14cm}
      \caption[]{Left panel: zoom of the first part of the GRO J1655-40 outburst as seen by BAT (blue) and HEXTE (red) (symbols as in Figure \ref{fig_GRO_flux}). The right panel represents the PCA count rate and hardness ratio as obtained from the ({\em http://tahti.mit.edu/opensource/1655/)} web site. The green box highlights the time span covered in the left panel.}
      \label{fig_GRO_flux53415_53470}
\end{center}
\end{figure*}

\begin{figure*}[htbp]
\begin{center}
       \hspace{0cm}\psfig{figure=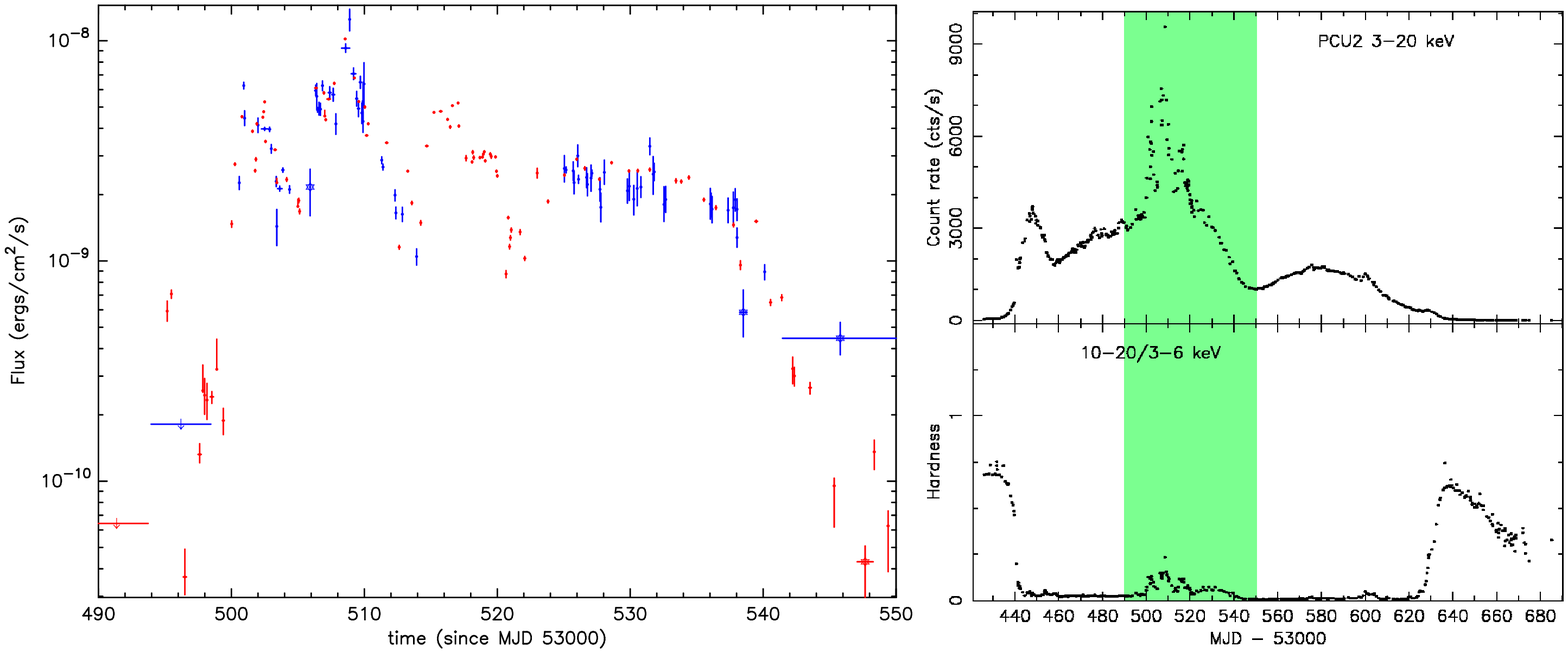,width=14cm}
      \caption[]{Zoom of the central part of the GRO J1655-40 outburst as seen by BAT (blue) and HEXTE (red).}
      \label{fig_GRO_flux53490_53550}
\end{center}
\end{figure*}

\begin{figure*}[htbp]
\begin{center}
       \hspace{0cm}\psfig{figure=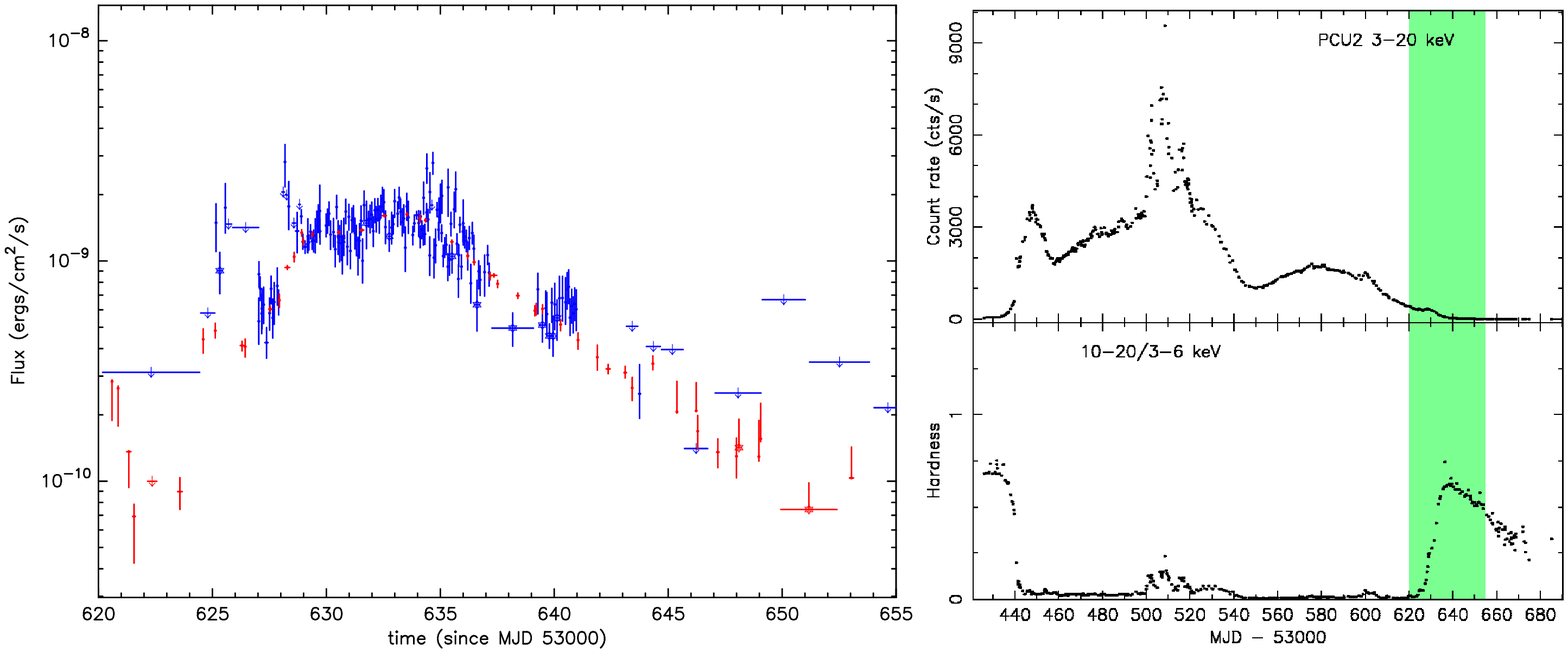,width=14cm}
      \caption[]{Zoom of the final part of the GRO J1655-40 outburst as seen by BAT (blue) and HEXTE (red).}
      \label{fig_GRO_flux53620_53655}
\end{center}
\end{figure*}

Zooms of sections of the light-curve are shown in figures
\ref{fig_GRO_flux53415_53470}, \ref{fig_GRO_flux53490_53550}  and
\ref{fig_GRO_flux53620_53655}, where the light-curves and hardness
ratio plots obtained with the PCA instrument (taken from {\em
http://tahti.mit.edu/opensource/1655/})~are also given. In spite of
the different time coverage of such a strongly variable source,
the agreement between the BAT and HEXTE light curves is
remarkably good. 

It is somewhat difficult to perform a
direct quantitative comparison since BAT and HEXTE observations
are not strictly simultaneous and the source shows a 
large variability on short timescales.
Generally, BAT and HEXTE measurements appear
to be fully consistent within errors.
Considering time windows for which the BAT and 
HEXTE observations are frequent and close in time, 
a difference not larger than
$\sim$10-15\% is apparent when the source flux is above 
1-2 $\times$10$^{-9}$ erg cm$^{-2}$s$^{-1}$, or $\sim90$ mCrab. Generally, a good agreement (within
errors) is found when the S/N in BAT
spectra is greater than 4. The actual
flux yielding such a S/N obviously depends on the position
of the target within the FOV.
Indeed, in one hour exposure, the $3\sigma$ sensitivity
with our approach is $\sim10-20$ mCrab for
an on-axis source,
while it is a factor $\sim10$ worse at
a coded fraction 0.2.
Thus, if the target lies within the half-coded
region, our approach
yields significant spectral
measurements (consistent with HEXTE)
in the 30-100 keV range
down to $(5-6) \times
10^{-10}$ erg cm$^{-2}$s$^{-1}$,
or $\sim50$ mCrab
(see e.g. Fig \ref{fig_GRO_flux53620_53655}, around MJD 53640). The study of sources fainter than $\sim50$ mCrab
would require a different and more complex approach.

\begin{figure}[htbp]
\begin{center}
       \hspace{0cm}\psfig{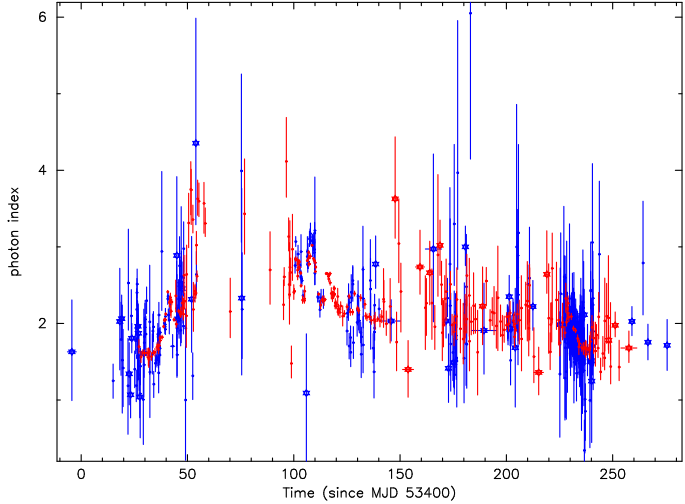}
      \caption[]{Spectral photon index evolution of the GRO J1655-40 outburst as obtained by fitting a power law independently to BAT (blue) and HEXTE (red) spectra. Errors are given at 1$\sigma$ level. Star labeled BAT and HEXTE points correspond to averaged spectra.}
      \label{fig_GRO_phindx}
\end{center}
\end{figure}

\begin{figure}[htbp]
\begin{center}
       \hspace{0cm}\psfig{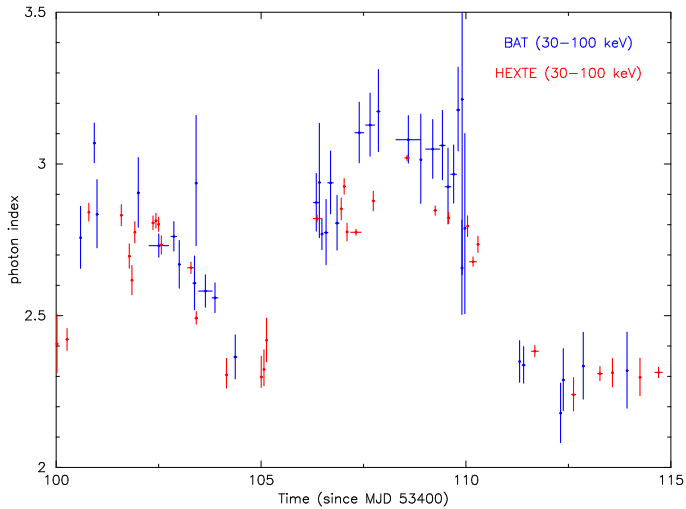}
      \caption[]{Zoom of figure \ref{fig_GRO_phindx} for the central part of the outburst.}
      \label{fig_GRO_phindx_sel}
\end{center}
\end{figure}

A good agreement between the power law photon index values as
measured by BAT and HEXTE is also apparent in
Figure~\ref{fig_GRO_phindx}. This is particularly evident if we
consider the time intervals corresponding to the highest source
flux (above $\sim$2 $\times$ 10$^{-9}$ ergs cm$^{-2}$ s$^{-1}$),
as shown in Fig. \ref{fig_GRO_phindx_sel}, where BAT and HEXTE
values agree to within $\sim$13\% with no apparent correlation
between spectral shape and flux discrepancy.

Fig.~\ref{fig_GRO_flux} is a clear proof of the 
good capabilities of our method 
in order to extract flux and spectral informations
for bright hard X-ray sources. BAT serendipitous coverage yielded a monitoring with a time
coverage fully comparable to that obtained through a systematic campaign with RXTE.
We note that BAT, simply owing to its good sensitivity over a very large FOV, 
caught the outburst since the very beginning, while the detection of
the source activity by HEXTE and PCA was due to planned observational campaign of 
the Galactic center region (PCA Galactic bulge scans).
Had GRO J1655-40 be located outside the galactic bulge region scanned by RXTE, its outburst 
would have been detected by ASM with $>$15 day delay with respect to BAT (see figure \ref{fig_GRO_flux53415_53470}).

\section{Summary and conclusions}\label{Discussion}

We have 
developed and tested a procedure,
based on public available Swift software tools, aimed at extracting
flux and spectral information for bright hard X-ray sources from
BAT ``survey'' data.
Tests performed using a large sample of Crab data
have shown that our pipeline, based  on the mask-weighting technique,
yields reliable measurements, both in spectral shape and
in flux, over the whole BAT FOV.
Using those results as a starting point, a more comprehensive test of
our method was carried out on a fainter, strongly
variable source such as GRO J1655-40. Its 9-month long outburst
was systematically monitored with PCA and HEXTE instruments on
board RXTE satellite and serendipitously observed by BAT. The
cross-check performed by analysing indipendently the BAT and
HEXTE spectra has shown a very good agreement between the two
instruments when the source signal-to-noise in BAT spectrum is greater than 4 
($\sim$50 mCrab, for a target within the half coded region),
confirming the reliability of our approach 
for using BAT as a hard X-ray
monitor for bright sources. 
Combining such good performances with the huge BAT FOV, which
covers each day $\sim$50\% of the sky, one realizes the
instrument's potential to frequently monitor the temporal and
spectral evolution of numerous bright, hard-X ray sources. As
shown in section \ref{GRO_results},
BAT has been able to detect the beginning of the GRO J1655-40
outburst almost simultaneously with PCA instrument, which was
luckily on target. Thus, while scanning the sky waiting for GRBs,
BAT can be used both for detecting emission from hard X-ray
transients and for monitoring the temporal and spectral evolution
of known sources.

\begin{acknowledgements}

We acknoledge useful comments from the BAT team.
We thank Gerry Skinner for useful discussion and support for data
analysis. ADL aknowledges an ASI fellowship. This work is
supported by ASI grant I/R/039/04 and PRIN 2005025417.

\end{acknowledgements}

\bibliographystyle{aa}
\bibliography{biblio_ARTICOLO}

\begin{thebibliography}{9}
\expandafter\ifx\csname natexlab\endcsname\relax\def\natexlab#1{#1}\fi

\bibitem[{{Barthelmy} {et~al.}(2005){Barthelmy}, {Barbier}, {Cummings},
  {Fenimore}, {Gehrels}, {Hullinger}, {Krimm}, {Markwardt}, {Palmer},
  {Parsons}, {Sato}, {Suzuki}, {Takahashi}, {Tashiro}, \&
  {Tueller}}]{Barthelmy}
{Barthelmy}, S.~D., {Barbier}, L.~M., {Cummings}, J.~R., {et~al.} 2005, Space
  Science Reviews, 120, 143

\bibitem[{{Gehrels} {et~al.}(2004){Gehrels}, {Chincarini}, {Giommi}, {Mason},
  {Nousek}, {Wells}, {White}, {Barthelmy}, {Burrows}, {Cominsky}, {Hurley},
  {Marshall}, {M{\'e}sz{\'a}ros}, {Roming}, {Angelini}, {Barbier}, {Belloni},
  {Campana}, {Caraveo}, {Chester}, {Citterio}, {Cline}, {Cropper}, {Cummings},
  {Dean}, {Feigelson}, {Fenimore}, {Frail}, {Fruchter}, {Garmire}, {Gendreau},
  {Ghisellini}, {Greiner}, {Hill}, {Hunsberger}, {Krimm}, {Kulkarni}, {Kumar},
  {Lebrun}, {Lloyd-Ronning}, {Markwardt}, {Mattson}, {Mushotzky}, {Norris},
  {Osborne}, {Paczynski}, {Palmer}, {Park}, {Parsons}, {Paul}, {Rees},
  {Reynolds}, {Rhoads}, {Sasseen}, {Schaefer}, {Short}, {Smale}, {Smith},
  {Stella}, {Tagliaferri}, {Takahashi}, {Tashiro}, {Townsley}, {Tueller},
  {Turner}, {Vietri}, {Voges}, {Ward}, {Willingale}, {Zerbi}, \&
  {Zhang}}]{Gehrels}
{Gehrels}, N., {Chincarini}, G., {Giommi}, P., {et~al.} 2004, \apj, 611, 1005

\bibitem[{{Hynes} {et~al.}(1998){Hynes}, {Haswell}, {Shrader}, {Chen}, {Horne},
  {Harlaftis}, {O'Brien}, {Hellier}, \& {Fender}}]{Hynes_1998}
{Hynes}, R.~I., {Haswell}, C.~A., {Shrader}, C.~R., {et~al.} 1998, \mnras, 300,
  64

\bibitem[{{Krimm} {et~al.}(2006){Krimm}, {Barbier}, {Barthelmy}, {Cummings},
  {Fenimore}, {Gehrels}, {Markwardt}, {Palmer}, {Parsons}, {Sakamoto}, {Sato},
  {Sanwal}, {Stamatikos}, \& {Tueller}}]{Krimm_2006}
{Krimm}, H., {Barbier}, L., {Barthelmy}, S.~D., {et~al.} 2006, The Astronomer's
  Telegram, 904, 1

\bibitem[{{Levine} {et~al.}(1984){Levine}, {Lang}, {Lewin}, {Primini},
  {Dobson}, {Doty}, {Hoffman}, {Howe}, {Scheepmaker}, {Wheaton}, {Matteson},
  {Baity}, {Gruber}, {Knight}, {Nolan}, {Pelling}, {Rothschild}, \&
  {Peterson}}]{Levine_1984}
{Levine}, A.~M., {Lang}, F.~L., {Lewin}, W.~H.~G., {et~al.} 1984, \apjs, 54,
  581

\bibitem[{{Markwardt} \& {Swank}(2005)}]{Markwardt_2005_GRO}
{Markwardt}, C.~B. \& {Swank}, J.~H. 2005, The Astronomer's Telegram, 414, 1

\bibitem[{{Markwardt} {et~al.}(2005){Markwardt}, {Tueller}, {Skinner},
  {Gehrels}, {Barthelmy}, \& {Mushotzky}}]{Markwardt_2005}
{Markwardt}, C.~B., {Tueller}, J., {Skinner}, G.~K., {et~al.} 2005, \apjl, 633,
  L77

\bibitem[{{Ulrich-Demoulin} \& {Molendi}(1996)}]{Molendi_1996}
{Ulrich-Demoulin}, M.-H. \& {Molendi}, S. 1996, \apj, 457, 77

\bibitem[{{Zhang} {et~al.}(1997){Zhang}, {Ebisawa}, {Sunyaev}, {Ueda},
  {Harmon}, {Sazonov}, {Fishman}, {Inoue}, {Paciesas}, \&
  {Takahashi}}]{Zhang_1997}
{Zhang}, S.~N., {Ebisawa}, K., {Sunyaev}, R., {et~al.} 1997, \apj, 479, 381

\end{thebibliography}

\begin{appendix}

\section{Data analysis pipelines}\label{Crab_pipeline}
In this section we provide a detailed description of the automatic
pipelines used throughout this work. We divide the pipeline into
different sections: 1) Preliminary data selection and
preparation; 2) Imaging analysis; 3) mask
weighting;
 4) spectral analysis.
\subsection{Preliminary data selection and preparation.}\label{Preliminary_data_selection_and_preparation}
We describe here the prescription we used to select
a good-quality, reliable dataset.
\begin{itemize}
\item First of all, we select data, collected in
the desired time interval, with the target inside the BAT FOV.
Data, including housekeeping and auxiliary files, may be searched
and retrieved through the \Swift~data archive (see {\em
http://heasarc.gsfc.nasa.gov/cgi-bin/W3Browse/swift.pl}).
\item We correct DPH energy scale using gain offset maps. Gain offset files,
stored among housekeeping files, are not supplied for each obs id.
The most recent available file is selected for each
observation, (the mean time between the file and the next is $\sim$3h).
The ad-hoc tool {\em baterebin} is used to perform the correction.
\item We reject bad datasets. Data filtering is performed using information
stored both in housekeeping and attitude files:
\begin{itemize}
\item Attitude information. Data for which the star
tracker is not locked and the spacecraft is not in pointing mode
are rejected. 
\item Number of enabled detectors. To optimize imaging capabilities and
detector performances, a minimum number of enabled detectors is required.
DPH collected with less than 24000 enabled detectors are rejected.
\item Background noise. DPHs with anomalously high background noise may be identified
by checking the total detector count rate. If such
value exceeds 18000 counts s$^{-1}$ in the 14-190 keV energy band, data are discarded.
\item Avoidance angles and occultations. Data for which the
angle between pointing and Earth limb is less than 30 degrees, or the spacecraft is in
South Atlantic Anomaly (SAA) are discarded. Time intervals during which the
source under study is occulted by Earth, Moon and Sun may be identified using
the ad-hoc {\em batoccultgti} tool.
\item Overall data quality. DPHs for which data quality is nominal may be
  identified by checking appropriate flag stored in each DPH file. 
\end{itemize}
Each of the above criteria yields a Good Time Interval (GTI)
table. The intersection of all such GTIs yields a table {\bf we} used to
screen for bad data. The resulting GTI is possibly unnecessarily
restrictive. Often good data may be discarded because of a few
seconds bad GTI intersection. In order to avoid such a problem, we added on both sides of each time interval a
small allowance of 10\% of the time interval. If such a value exceeds 10 seconds, we set the allowance to 10 s.
Next, we checked all DPH times
against resulting GTIs using the {\em batbinevt} tool, both for entire DPH and for a combination of DPH rows. Only the total intersected data will be
considered for further analysis. 
\item We update attitude information.
A new attitude file based on the median pointing direction, as evaluated
considering only the observation GTI,
is generated using the {\em aspect} tool.
\end{itemize}

\subsection{Imaging analysis and source detection.}\label{Imaging_Analysis}
We give here details on 
the procedure we used to
extract
an energy resolved image of the sky and to search
for the source of interest as well as for new ones.
\begin{itemize}
\item we create a Detector Plane Image (DPI). DPIs are obtained
from DPH by adding, for each detector, the total counts recorded.
Energy selections may be performed a priori in order to have energy-resolved
DPIs. Such operations may be performed using the {\em batbinevt} tool.
\item we generate a detector mask starting from the map of enabled/disabled
detectors stored in the housekeeping files.
As for the case of gain offset files, detector masks are not supplied for each
observation.
They are supposed not to change rapidly, so
we used the nearest available file.
Hot pixels are then searched and identified using the ad-hoc task
{\em bathotpix}. A bad pixel map is obtained, and it is combined
to the original map to produce the final detector mask.
\item  we create a background map for each DPI, not accounting
  for sources inside the FOV, using the {\em batclean} tool, the detector mask, and the default background model.
\item we built the sky image. Each DPI was deconvolved from the
coded aperture mask and translated to a background-subtracted
image of the sky (including all the
sources within the instrument FOV) with the ad-hoc task {\em batfftimage}, using
the background map, the corrected attitude file and the merged detector mask.
\item  we create a coded fraction map. A map of the sky coded fraction across the
  instrument FOV was obtained using the {\em batfftimage} tool using the
same inputs as above. Values of 1 are
applied to fully coded regions, 0 to the edges of FOV.
\item we run the source detection algorithm. We used the ad-hoc task {\em batcelldetect} 
to perform a source detection on the resulting sky images. A list
of sources above a desired signal-to-noise threshold (we adopted 3.8) and a partially-coded fraction threshold (we adopted 0.001) is
generated, including sky and image coordinates, signal-to-noise,
count rate, coded fraction and other useful pieces of information
for each source.

\end{itemize}

The procedure outlined above may be run either on a single-row DPH, or
on a merged DPH.

\subsection{Mask-weighting technique.}\label{Mask_weighting}

We describe here the procedure we used to extract the source spectrum 
using the mask weighting approach,
as well as to generate appropriate response matrix. The source coordinates 
must be known a priori (either from source detection,
or from independent measurements).
This section of our
pipeline uses as input either a single-row DPH, or a merged DPH. 

\begin{itemize}
\item We generate the weight map for the source under study for each DPH.
The ad-hoc tool {\em batmaskwtimg} is used to this aim, using
the detector mask as well as the corrected  attitude files described above.
\item We extract the source background-subtracted spectrum
from each DPH. This is done
with the {\em batbinevt} tool, using
energy corrected, GTI-filtered DPH together with the corresponding
weight map.
\item We generate an ad-hoc response matrix response matrix (including effective area
information) for each spectrum using the task {\em batdrmgen}.
\item We add systematic error.
The spectrum extracted as described above does not include systematic errors,
which may be relevant
at low energy ($<$25 keV). The {\em batphasyserr} tool
may be used to add ad-hoc (as stored in the CALDB) systematics to each spectrum.
\end{itemize}

\subsection{Spectral Analysis.}
This section of the pipeline 
is devoted to the automatic spectral analysis
of a source strongly variable both in flux and in spectral shape.
We proceed as follows:
\begin{itemize}
\item Evaluation of the target signal-to-noise. \\
As a first step, we estimate the source signal-to-noise for each
spectrum. This is simply done by selecting a suitable
spectral range, which has to be optimized on a case by case basis, and checking the source count rate together with its associated error.
If the ratio between the count rate and the error is null or negative,
the spectrum (and the corresponding DPH group) is discarded. If such ratio
is positive, then further steps are performed.
\item Initial guess of the spectral shape using hardness ratio {which is required } in order to optimize an automatic, ``blind'' spectral analysis.
For the case of a strongly variable source, we used a hardness ratio
criterion. Such an approach may allow to discriminate
between different source states, and/or to identify the presence
and relative contribution of different spectral components. The
resulting hardness ratio values allow to choose the spectral
model to be used to fit the source spectra, and/or to select an
appropriate energy range to perform spectral analysis.
GRO J1655-40 shows a strong, variable thermal component, dominating the
source spectrum below 10 keV, while up to $\sim100$ keV the spectrum
is well reproduced by a power law (\citep{Zhang_1997};~\citep{Hynes_1998}).
 Sometimes the thermal component yields a
very significant (or even dominant) contribution up to more than
20 keV. In order to simplify the spectral analysis using a
single-component model, the hardness ratio test was used to select
the energy range minimizing the thermal contribution.
In particular, we considered the
hardness ratio between the 16-22 keV and 20-70 keV
energy ranges. Spectra with a high count rate in the soft band are studied
in 30-100 keV energy range, otherwise the 10-100 keV band is used.
\item Evaluation of the source flux. \\
Before running a complex spectral fit (also in view of the
hardness ratio results), we perform a preliminary
fit with a simplified model (reducing, e.g., the number of free
spectral parameters) to get the source flux with its associated
error. We used the flux/error ratio as a criterium to decide
if the data quality warrant a deeper spectral model. In particular,
a preliminary power law fit with a photon index fixed to -3 was
performed and the source flux was evaluated. A flux/error ratio
threshold of 4 was used to discriminate between low and high
S/N spectra.
\item Low S/N spectra. \\ Spectra with poor S/N are used to set an upper limit
to the source emission using the simplified model.
\item Further analysis of low S/N spectra. \\
We made an attempt to recover spectral information, by summing low
S/N spectra extracted from consecutive groups of DPHs, up to a
maximum of 10
contiguous spectra. The corresponding response matrices are also
combined. The resulting spectra are then analysed through the
same steps of the automated pipeline.
\item High S/N spectra. \\
Spectra with an adequate S/N are studied in detail within Xspec.
After spectral fitting, the best spectral parameters, as well as
the source refined flux, are computed together with their
uncertainties. In our case, a power law was used. The
normalization was evaluated at 40 keV in order to minimize
correlation with the photon index \citep[see][]{Molendi_1996}.

\end{itemize}

\section{Attitude information for stacked DPHs}\label{Aspect}
In order to understand the section dealing with the stacking of different DPH,
focused on the maximum allowed misalignement between the observations,
a brief explanation about the {\em aspect} tool,
handling the attitude information on DPHs, is necessary. {\em aspect}
calculates the mean pointing for a given attitude file as follows: the algorithm first creates a two-dimensional
histogram of the amount of time spent at each RA and DEC, then selects the bin with the largest time
and calculates the mean RA, DEC and ROLL angle of the spacecraft axes while RA and DEC angles were in this bin.
The default binsize for such an histogram is 0.01 degs 
and usually a large fraction of the integration time of a DPH refers to 
a single bin of such size.
This means that, if we stack two DPHs with an offset larger than 0.01 degs,
the resulting pointing coordinates as obtained with {\em aspect} will be the same of
the fraction of DPH 
accounting for the peak in the above described histogram.
In the case of two histogram bins with the same amount of time, aspect calculates the mean of the relevant coordinates.
If the offset is smaller than 0.01 degs (or than the selected binsize), {\em aspect} calculates
the mean of the coordinates, as described above.
Our approach was to leave fixed the default {\em aspect} binsize value and to choose,
for each of the three coded fraction regions selected above, a reference single DPH's row with 450 s exposure time: any other selected DPH's row with the same exposure time was stacked on it.

In practice, considering offset up to few arcmin, the use of a larger {\em aspect} binsize
does not change significantly the flux estimation of the source.

\end{appendix}

\end{document}